\documentclass[aps,prd,nofootinbib,twocolumn,superscriptaddress,preprintnumbers,floatfix]{revtex4-1}
\usepackage{graphicx}
\usepackage{amsmath} 
\usepackage[T1]{fontenc} 
\usepackage{amssymb}
\usepackage[usenames,dvipsnames]{xcolor}
\usepackage{cleveref}
\usepackage{upgreek}
\usepackage{listings}
\usepackage{lipsum}
\usepackage{slashed}
\usepackage{mathtools}
\usepackage{ctable}
\usepackage{feynmp-auto}

\newcommand{\prn}[1]{ \left(  #1 \right) }

\newcommand{\avg}[1]{\left< #1 \right>}

\newcommand{\magn}[1]{\left| #1 \right|}

\newcommand{\al}[1]{\begin{align} #1 \end{align}}

\newcommand{\half}{\frac{1}{2}}
\newcommand{\Neff}{N_{\text{eff}}}
\newcommand{\gamp}{{\gamma '}}
\newcommand{\QCD}{\Lambda_{\text{QCD}}}
\newcommand{\dd}{\partial}

\newcommand{\ep}{{e'}}
\newcommand{\ebp}{{\bar{e}'}}
\newcommand{\oo}{\infty}
\newcommand{\MM}{\mathcal{M}}

\begin{document}
\title{Freeze-\textit{Tw}in Dark Matter}

\author{Seth Koren}
\affiliation{Department of Physics, University of California, Santa Barbara, CA 93106, USA}
\author{Robert McGehee}
\affiliation{Berkeley Center for Theoretical Physics, University of California, Berkeley, CA 94720}
\affiliation{Theoretical Physics Group, Lawrence Berkeley National Laboratory, Berkeley, CA 94720}

\begin{abstract}
\noindent
The mirror twin Higgs (MTH) addresses the little hierarchy problem by relating every Standard Model (SM) particle to a twin copy, but is in tension with cosmological bounds on light degrees of freedom. Asymmetric reheating has recently been proposed as a simple way to fix MTH cosmology by diluting the twin energy density. We show that this dilution sets the stage for an interesting freeze-in scenario where both the initial absence of dark sector energy and the feeble coupling to the SM are motivated for reasons unrelated to dark matter production. We give the twin photon a Stueckelberg mass and freeze-in twin electron and positron dark matter through the kinetic mixing portal. The kinetic mixing required to obtain the dark matter abundance is of the loop-suppressed order expected from infrared contributions in the MTH.
\end{abstract}

\maketitle
\section{Introduction}

The electroweak hierarchy problem has long provided theoretical motivation for the presence of new TeV-scale physics. 
With the curious absence of new physics observed at the LHC, the issue of Higgs naturalness becomes increasingly urgent and prompts the consideration of new approaches to this old problem. 
One such approach is the `Neutral Naturalness' paradigm in which the states responsible for stabilizing the electroweak scale are not charged under (some of) the Standard Model (SM) gauge symmetries \cite{Chacko:2005pe,Chacko:2005un,Burdman:2006tz,Poland:2008ev,Cai:2008au,Craig:2014aea,Batell:2015aha,Curtin:2015bka,Cheng:2015buv,Craig:2016kue,Cohen:2018mgv,Cheng:2018gvu,Serra:2019omd}, thus explaining the lack of expected signposts of naturalness. 

The first and perhaps most aesthetically pleasing of these Neutral Naturalness approaches is the mirror twin Higgs (MTH) \cite{Chacko:2005pe}, which 
stabilizes the Higgs potential up to a cutoff $\Lambda \sim 5-10\text{ TeV}$. In the MTH, the states responsible for ensuring naturalness comprise 
another copy of the SM related to ours by a $\mathbb{Z}_2$ symmetry. 
Since these `twin' states are 
neutral under the SM gauge group, they easily evade LHC constraints.
The most serious empirical challenges to these models comes instead from cosmology. In particular, the presence of 
thermalized light twin neutrinos and photons~\cite{Craig:2016lyx} is significantly ruled out by BBN~\cite{Cyburt:2015mya} and the CMB~\cite{Aghanim:2018eyx}.

This cosmological tension has been alleviated with the observation that an exact $\mathbb{Z}_2$ symmetry is not truly necessary for naturalness.
This was first exploited in \cite{Craig:2015pha}, where only the third generation of SM fermions was copied and twin hypercharge was not gauged. This philosophy has since been taken up by the authors of \cite{Barbieri:2016zxn,Csaki:2017spo,Liu:2019ixm,Harigaya:2019shz} who  
 introduce clever $\mathbb{Z}_2$-asymmetries  
into the MTH mass spectrum  
to alleviate cosmological problems. 
These models 
have proved to be a boon for phenomenology. Among other things, they
quite generally motivate looking for 
Higgs decays to long-lived particles at colliders \cite{Clarke:2015ala,Curtin:2015fna,Csaki:2015fba,Pierce:2017taw,Alipour-Fard:2018lsf,Kilic:2018sew,Alipour-fard:2018mre,Li:2019ulz} 
and contain well motivated dark matter (DM) candidates~\cite{Garcia:2015loa,Craig:2015xla,Garcia:2015toa,Farina:2015uea,Freytsis:2016dgf,Farina:2016ndq,Prilepina:2016rlq,Barbieri:2017opf,Hochberg:2018vdo,Cheng:2018vaj,Terning:2019hgj,Earl:2019wjw}. 

In this work, we build instead on a framework where `hard' breaking of the $\mathbb{Z}_2$ is absent. In \cite{Craig:2016lyx,Chacko:2016hvu}, it was realized that late-time asymmetric reheating of the two sectors could arise naturally in these models if the spectrum were extended by a single new state.
This asymmetric reheating would dilute the twin energy density and so attune the MTH with the cosmological constraints. 
This dilution of twin energy density to negligible levels would seem to hamper the prospect that twin states might constitute the dark matter, and generating dark matter was left as an open question. This presents a major challenge toward making such cosmologies realistic. 
However, we show that asymmetric reheating perfectly sets the stage for a MTH realization of the `freeze-in' mechanism for dark matter production~\cite{Asaka:2005cn,Gopalakrishna:2006kr,Asaka:2006fs,Page:2007sh,Hall:2009bx,Chu:2011be,Bernal:2017kxu,Dvorkin:2019zdi}.

Freeze-in scenarios are characterized by two assumptions: 1) DM has a negligible density at some early time and 2) DM interacts with the SM so feebly that it never achieves thermal equilibrium with the SM bath.\footnote{We note that the feeble connection between the two sectors may originate as a small dimensionless coupling or as a small ratio of mass scales, either of which deserves some explanation.} This second assumption is motivated in part by the continued non-observation of non-gravitational DM-SM interactions.  
Both assumptions stand in stark contrast to freeze-out scenarios. 

Freeze-twin dark matter is a particularly interesting freeze-in scenario because both assumptions are fulfilled for reasons orthogonal to dark matter considerations: 1) the negligible initial dark matter abundance is 
predicted by the asymmetric reheating already necessary to resolve the MTH cosmology, and 2) the kinetic mixing necessary to achieve the correct relic abundance is of the order expected from infrared contributions in the MTH. 
To allow the frozen-in twin electrons and positrons to be DM, we need only 
break the $\mathbb{Z}_2$ by a relevant operator to give a Stueckelberg mass to twin hypercharge.
Additionally, the twin photon masses we consider can lead to dark matter self-interactions at the level relevant for small-scale structure problems~\cite{Tulin:2017ara}. 

This paper is structured as follows. In Section \ref{sec:mth}, we review the MTH and 
its cosmology in models with asymmetric reheating, and
in Section \ref{sec:kin} we introduce our extension. 
In Section \ref{sec:freezein}, we calculate the freeze-in yield for twin electrons and discuss the parameter space to generate dark matter and constraints thereon. 
We discuss future directions and conclude in Section \ref{sec:conc}. For the interested reader, we include some discussion of the irreducible IR contributions to kinetic mixing in the MTH in Appendix \ref{sec:KinMixapp}.

\section{The Mirror Twin Higgs \& Cosmology}\label{sec:mth}

The mirror twin Higgs  
framework \cite{Chacko:2005pe} introduces a twin sector $B$, which is a `mirror' copy of the Standard Model sector $A$, related by a $\mathbb{Z}_2$ symmetry. 
Upgrading the $SU(2)_{A}\times SU(2)_{B}$ gauge symmetry of the scalar potential to an $SU(4)$ global symmetry adds a Higgs-portal interaction between the $A$ and $B$ sectors:
\begin{equation}
\label{eq:MTHV}
V = \lambda \left( |\mathcal{H}|^2 - f^2/2\right)^2,
\end{equation}
where $\mathcal{H} = \begin{pmatrix} H_A \\ H_B \end{pmatrix}$ is a complex $SU(4)$ fundamental consisting of the $A$ and $B$ sector Higgses in the gauge basis. The SM Higgs is to be identified as 
a pseudo-Goldstone mode arising from the breaking of $SU(4)\rightarrow SU(3)$ when $\mathcal{H}$ acquires a vacuum expectation value (vev) $\langle \mathcal{H} \rangle = f/\sqrt{2}$. Despite the fact that the global $SU(4)$ is explicitly broken by the gauging of $SU(2)_{A}\times SU(2)_{B}$ subgroups, the $\mathbb{Z}_2$ is enough to ensure that the quadratically divergent part of the one-loop effective action respects the full $SU(4)$.  The lightness of the SM Higgs is then understood as being protected by the approximate accidental global symmetry up to the UV cutoff scale $\Lambda \lesssim 4 \pi f$, at which point new physics must come in to stabilize the scale $f$ itself.

Current measurements of the Higgs couplings at the LHC are consistent with the SM, implying  $f \gtrsim 3v$~\cite{Burdman:2014zta,Craig:2015pha}, where $v$ is the SM Higgs vev.
This requires some vacuum misalignment via soft $\mathbb{Z}_2$-breaking which leads the two $SU(2)$ doublets to get vevs $v_A \approx v/\sqrt{2}$ and $v_B \approx f/\sqrt{2}$, to lowest order. The twin spectrum is thus 
a factor of $f/v$ heavier than the standard model spectrum. 
We refer to twin particles by their SM counterparts primed with a superscript ', and we refer the reader to \cite{Chacko:2005pe,Craig:2015pha} for further discussion of the twin Higgs mechanism.

The thermal bath history in the conventional MTH is fully dictated by the Higgs portal in Eq.~\eqref{eq:MTHV}  
which keeps the SM and twin sectors in thermal equilibrium down to temperatures $\mathcal{O}(\text{GeV})$. 
A detailed calculation of the decoupling process was performed in \cite{Craig:2016lyx} by tracking the bulk heat flow between the two sectors as a function of SM temperature. It was found that for the benchmark of $f/v=4$, decoupling begins at a SM temperature of $T \sim 4 \text{ GeV}$ and by $\sim 1 \text{ GeV}$, 
the ratio of twin-to-SM temperatures may reach $\lesssim 0.1$  without rebounding.
While heat flow rates become less precise below $\sim 1 \text{ GeV}$ due to uncertainties in hadronic scattering rates, especially close to color-confinement, decoupling between the two sectors is complete by then for $f/v \gtrsim 4$.
For larger $f/v$, the decoupling begins and ends at higher temperatures. 

After the two sectors decouple, chemical processes in the two sectors change their temperatures independently. 
The twin sector eventually cools to a slightly lower temperature than the SM due to 
the modification of mass thresholds. 
However, 
within a standard cosmology, this still leaves far too much radiation in the twin sector to agree with BBN and CMB observations. 
To quantify this tension, the effective number of neutrinos, $\Neff$, is defined in
\al{
\rho_r = \prn{1+\frac{7}{8} \prn{\frac{4}{11}}^{4/3} \Neff} \rho_\gamma,
}
where $\rho_r$ is the total radiation energy density, the factors $7/8$ and $\prn{4/11}^{4/3}$ come from Fermi-Dirac statistics and SM electron-positron annihilations, and $\rho_\gamma$ is the SM photon energy density. The SM neutrinos contribute $\Neff \approx 3.046$~\cite{deSalas:2016ztq,Mangano:2001iu}, with additional radiative degrees of freedom collectively contributing to $\Delta \Neff \equiv \Neff - 3.046$. 
In a generic MTH model, the twin neutrinos and photon contribute $\Delta \Neff \sim 5-6$~\cite{Chacko:2016hvu,Craig:2016lyx}, significantly disfavored by both BBN~\cite{Cyburt:2015mya} and the more stringent Planck measurement, $2.99^{+0.34}_{-0.33}$ (at 95\% confidence)~\cite{Aghanim:2018eyx}.\footnote{
Care must be taken  
when applying this constraint since the twin neutrinos may be semi-relativistic 
by matter-radiation equality~\cite{Craig:2016lyx}. But, within a standard cosmology, the MTH is unambiguously ruled out.}

As mentioned above, one class of solutions to this $\Neff$ problem uses hard breaking of the $\mathbb{Z}_2$ at the level of the spectra~\cite{Craig:2015pha,Barbieri:2016zxn,Csaki:2017spo,Liu:2019ixm,Harigaya:2019shz} while keeping a standard cosmology. An alternative proposal is to
modify the cosmology with asymmetric reheating to dilute the energy density of twin states. 
For example, \cite{Chacko:2016hvu} uses late, out-of-equilibrium decays of right-handed neutrinos, while \cite{Craig:2016lyx} uses those of a scalar singlet.
These new particles respect the $\mathbb{Z}_2$, but dominantly decay to SM states due to the already-present soft $\mathbb{Z}_2$-breaking in the scalar sector.
In \cite{Chacko:2016hvu}, this is solely due to extra suppression by $f/v$-heavier mediators, 
while in \cite{Craig:2016lyx}, the scalar also preferentially mass-mixes with the heavier twin Higgs. \cite{Craig:2016lyx} also presented a toy model of `Twinflation', where a softly-broken $\mathbb{Z}_2$-symmetric scalar sector may lead to inflationary reheating which asymmetrically reheats the two sectors to different temperatures.
In any of these scenarios, the twin sector may be diluted to 
the level where it evades Planck bounds \cite{Aghanim:2018eyx} on extra radiation, yet is potentially observable with 
CMB Stage IV~\cite{Abazajian:2016yjj}. 

We will stay agnostic about the particular mechanism at play, and merely assume that by $T \sim 1 \text{ GeV}$, the Higgs portal interactions have become inefficient and some mechanism of asymmetric reheating has occurred such that the energy density in the twin sector has been largely depleted, $\rho_{\text{twin}} \approx 0$.\footnote{If asymmetric reheating leaves some small $\rho_{\text{twin}}>0$, then mirror baryon asymmetry can lead to twin baryons as a small subcomponent of dark matter \cite{Chacko:2018vss}.} This is consistent with the results of the decoupling calculation in \cite{Craig:2016lyx} given the uncertainties in the rates at low temperatures, and will certainly be true once one gets down to $\text{few } \times  10^2 \text{ MeV}$. 

One may be concerned that there will be vestigial model-dependence from irrelevant operators induced by the asymmetric reheating mechanism which connect the two sectors. However, these operators will generally be suppressed by scales above the reheating scale, as in the example studied in~\cite{Craig:2016lyx}. Prior to asymmetric reheating, the two sectors are in thermal equilibrium anyway, so these have little effect. After the energy density in twin states has been diluted relative to that in the SM states, the temperature is far below the heavy masses suppressing such irrelevant operators, and thus their effects are negligible. So we may indeed stay largely agnostic of the cosmological evolution before asymmetric reheating as well as the details of how this reheating takes place. We take the absence of twin energy density as an initial condition, but emphasize that there are external, well-motivated reasons for this to hold in twin Higgs models, as well as concrete models that predict this occurrence naturally.

\section{Kinetic Mixing \& a Massive Twin Photon} \label{sec:kin}

In order to arrange for freeze-in, we add to the MTH kinetic mixing between the SM and twin hypercharges and a Stueckelberg mass for twin hypercharge.  
At low energies, these reduce to such terms for the photons instead, parametrized as
\al{ \label{eq:addl}
\mathcal{L} \mathrel{+}= \frac{\epsilon}{2} F_{\mu\nu} F'^{\mu\nu} + \half m_{\gamma'}^2 A'_\mu A'^\mu.
}
This gives each SM particle of electric charge $Q$ an effective twin electric charge $\epsilon Q$.\footnote{Note that twin charged states do not couple to the SM photon. Their coupling to the SM Z boson has no impact on freeze-in at the temperatures under consideration. Furthermore, the miniscule kinetic mixing necessary for freeze-in has negligible effects at collider experiments. See Ref.~\cite{Chacko:2019jgi} for details.} The twin photon thus gives rise to a `kinetic mixing portal' through which the SM bath may freeze-in light twin fermions in the early universe.

The Stueckelberg mass constitutes soft $\mathbb{Z}_2$-breaking,\footnote{While we are breaking the $\mathbb{Z}_2$ symmetry by a relevant operator, the extent to which a Stueckelberg mass is truly \textit{soft} breaking is not clear. Taking solely Eq. (\ref{eq:addl}), we would have more degrees of freedom in the twin sector than in the SM, and in a given UV completion it may be difficult to isolate this $\mathbb{Z}_2$-asymmetry from the Higgs potential. 
One possible fix may be to add an extremely tiny, experimentally allowed Stueckelberg mass for the SM photon as well \cite{Goldhaber:2008xy}, though we note this may be in violation of quantum gravity \cite{Reece:2018zvv,Craig:2018yld} 
or simply be difficult to realize in UV completions without extreme fine-tuning. We will remain agnostic about this UV issue and continue to refer to this as `soft breaking', following \cite{Chacko:2019jgi}.} but has no implications for the fine-tuning of the Higgs mass since hypercharge corrections are already consistent with naturalness \cite{Craig:2015pha}. We will require $m_{\gamma'}> m_{e'}$, to prevent frozen-in twin electron/positron annihilations, and $m_{\gamma'}> 2 m_{e'}$, to ensure that resonant production through the twin photon is kinematically accessible.
Resonant production will allow a much smaller kinetic mixing to generate the correct relic abundance, thus avoiding indirect bounds from supernova cooling. We note that while taking $m_{\gamma'} \ll f$ does bear explanation, the parameter is technically natural. 

On the other hand, mixing of the twin and SM $U(1)$s preserves the symmetries of the MTH EFT, so quite generally one might expect it to be larger than that needed for freeze-in. However, it is known that in the MTH a nonzero $\epsilon$ is not generated through three loops \cite{Chacko:2005pe}. While such a suppressed mixing is phenomenologically irrelevant for most purposes, here it plays a central role. In Appendix \ref{sec:KinMixapp}, we discuss at some length the vanishing of infrared contributions to kinetic mixing through few loop order. If nonzero contributions appear at the first loop order where they are not known to vanish, kinetic mixing of the order $\epsilon \sim 10^{-13} - 10^{-10}$ is expected. 

The diagrams which generate kinetic mixing will likely also generate higher-dimensional operators. These will be suppressed by (twin) electroweak scales and so, as discussed above for the irrelevant operators generated by the model-dependent reheating mechanism, freeze-in contributions from these operators are negligible.

\section{Freezing-\textit{Tw}in Dark Matter} \label{sec:freezein}

As we are in the regime where freeze-in proceeds while the temperature sweeps over the mass scales in the problem, it is not precisely correct to categorize this into either ``UV freeze-in'' or ``IR freeze-in''. Above the mass of the twin photon, freeze-in proceeds through the marginal kinetic mixing operator, and so a naive classification would say this is IR dominated. However, below the mass of the twin photon, the clearest approach is to integrate out the twin photon, to find that freeze-in then proceeds through an irrelevant, dimension-six, four-Fermi operator which is suppressed by the twin photon mass. Thus, at temperatures $T_{\text{SM}} \lesssim m_{\gamma'}$, this freeze-in looks UV dominated. This leads to the conclusion that the freeze-in rate is largest at temperatures around the mass of the twin photon. Indeed, this is generally true of freeze-in --- production occurs mainly at temperatures near the largest relevant scale in the process, whether that be the largest mass out of the bath particles, mediator, and dark matter, or the starting thermal bath temperature itself in the case that one of the preceding masses is even higher.

As just argued, freeze-in production of dark matter occurs predominantly at and somewhat before $T \sim m_{\gamma'}$. 
We require $m_\gamp \ll 1 \text{ GeV}$ so that most of the freeze-in yield comes from when $T <1 \text{ GeV}$, which
allows us to retain `UV-independence' in that we need not care about how asymmetric reheating has occurred past providing negligible density of twin states at $T=1 \text{ GeV}$. Specifically, we limit ourselves to $m_{\gamma'} < 2 m_{\pi^0}$, both for this reason and to avoid uncertainties in the behavior of thermal pions during the epoch of the QCD phase transition. However, we emphasize that freeze-in will remain a viable option for producing a twin DM abundance for even heavier dark photons. But the fact that the freeze-in abundance will be generated simultaneously with asymmetric reheating demands that each sort of asymmetric reheating scenario must then be treated separately. Despite the additional difficulty involved in predicting the abundance for larger twin photon masses, it would be interesting to explore this part of parameter space. In particular, it would be interesting to consider concrete scenarios with twin photons in the range of tens of GeV \cite{Batell:2019ptb}.

To calculate the relic abundance of twin electrons and positrons, we use the Boltzmann equation for the number density of $e'$:
\al{
\dot{n}_\ep+3Hn_\ep=\sum_{k,l} -\avg{\sigma v}_{\ep \ebp \to kl} \prn{n_\ep n_\ebp-n_\ep^\text{eq} n_\ebp^\text{eq}},
}
where $\avg{\sigma v}_{\ep \ebp \to kl}$ is the thermally averaged cross section for the process $\ep \ebp \to kl$, the sum runs over all processes with SM particles in the final states and $\ep \ebp$ in the initial state, and $n_\ep^\text{eq}$ is the equilibrium number density evaluated at temperature $T$. As we are in the parametric regime in which resonant production of twin electrons through the twin photon is allowed, $2 \rightarrow 2$ annihilation processes $\bar{f} f \to \gamp \to \ebp \ep$, with $f$ a charged SM fermion, entirely dominate the yield. 

In accordance with the freeze-in mechanism, $n_\ep$ remains negligibly smaller than its equilibrium number density throughout the freeze-in process, and so that term is ignored. It is useful to reparametrize the abundance of $\ep$ in terms of its yield, $Y_\ep = n_\ep/s$ where $s=\frac{2\pi^2}{45} g_{\ast s}T^3$ is the entropy density in the SM bath. Integrating the Boltzmann equation using standard methods, we then find the yield of $\ep$ today to be
\al{
\label{eq:Yepexact}
Y_\ep&=\int_{0}^{T_i} dT \frac{\prn{45}^{3/2}}{\sqrt{2}\pi^3 \sqrt{g_\ast}g_{\ast s}} \frac{M_{Pl}}{T^5} \prn{\frac{1}{T}+\frac{\dd_T g_{\ast s}}{3 g_{\ast s}}} \nonumber \\
&\times \sum_{\bar{f}f \to \ebp \ep} \avg{\sigma v}_{\bar{f}f \to \ebp \ep} n_\ebp^\text{eq} n_\ep^\text{eq},
}
where $T_i = 1 \text{ GeV}$ is the initial temperature of the SM bath at which freeze-in begins in our setup, $g_\ast(T)$ is the number of degrees of freedom in the bath, and $M_{Pl}$ is the reduced Planck mass. We will calculate this to an intended accuracy of 50\%. To this level of accuracy, we may assume Maxwell-Boltzmann statistics to vastly simplify the calculation \cite{Blennow:2013jba}. As a further simplification, we observe that the $\partial_T g_{\star s}$ term is negligible compared to $1/T$ except possibly during the QCDPT - where uncertainties on its temperature dependence remain \cite{Drees:2015exa} - and so we ignore that term. 
The general expression for the thermally averaged cross section of the process $12 \to 34$ is then
\al{
\avg{\sigma v}n_1^\text{eq}n_2^\text{eq}&=\frac{T^4}{2^9 \pi^5 s_{34}} \int^\oo_{\text{Max} \prn{\frac{m_1+m_2}{T},\frac{m_3+m_4}{T}}} dx x^2 \\ 
&\times \sqrt{\left[1,2\right]} \sqrt{\left[3,4\right]} K_1 (x) \int d\prn{\cos \theta} \magn{\MM}^2_{12 \to 34}, \nonumber
}
where $s_{34}$ is 1 if the final states are distinct and 2 if not, $x=\sqrt{s}/T$, $\sqrt{\left[i,j\right]}=\sqrt{1-\prn{\frac{m_i+m_j}{xT}}^2}\sqrt{1-\prn{\frac{m_i-m_j}{xT}}^2}$, and $\magn{\MM}^2_{12 \to 34}$ is the matrix element squared summed (not averaged) over all degrees of freedom. 
To very good approximation, the yield results entirely from resonant production, and so we may analytically simplify the matrix element squared for $\bar{f}f \to \ebp \ep$ using the narrow-width approximation 
\al{
\int d\prn{\cos \theta} \magn{\MM}^2_{\bar{f}f \to \ebp \ep} &\approx \frac{256 \pi^3 \alpha^2 \epsilon^2}{3} \prn{2m_f^2+m_\gamp^2}  \\ 
&\times \frac{ \prn{2m_{e'}^2+m_\gamp^2}}{\Gamma_\gamp m_\gamp^2 T} \delta \prn{x-m_\gamp/T}. \nonumber
}
$\Gamma_\gamp$ is the total decay rate of the twin photon.

For the range of $m_\gamp$ we consider, the twin photon can only decay to twin electron and positron pairs. Thus, its total decay rate is 
\al{
\label{eq:Gamgamp}
\Gamma_{\gamp}=\frac{\alpha \prn{m_\gamp^2+2m_\ep^2}}{3m_\gamp} \sqrt{1-\frac{4m_\ep^2}{m_\gamp^2}}.
}
Its partial widths to SM fermion pairs are suppressed by $\epsilon^2$, and so contribute negligibly to its total width.

The final yield of twin electrons is then
\al{
\label{eq:Yep}
Y_\ep \! \approx \! \frac{3m_\gamp^2}{2\pi^2} \frac{\prn{45}^{3/2}M_{Pl}}{\sqrt{2} \pi^3} \! \sum_f \! \int_{T_f}^{T_i} \! dT \Gamma_{\gamp \to \bar{f} f} \frac{K_1(\frac{m_\gamp}{T})}{\sqrt{g_\ast} g_{\ast s}T^5}, \!
}
where $T_f=\QCD$ for quarks, $T_f=0$ for leptons, $\Gamma_{\gamp \to \bar{f} f}$ is the partial decay width of the twin photon to $f \bar{f}$, and the sum is over all SM fermion-antifermion pairs for which $m_\gamp > 2 m_f$.

Since we have approximated the yield as being due entirely to on-shell production and decay of twin photons, the analytical expression for the yield in Eq.~\eqref{eq:Yep} exactly agrees with the yield from freezing-in $\gamp$ via `inverse decays' $\bar{f}f \to \gamp$, as derived in~\cite{Hall:2009bx}. We have validated our numerical implementation of the freeze-in calculation by successfully reproducing the yield in similar cases found in \cite{Blennow:2013jba,Chang:2019xva}. We have furthermore checked that reprocessing of the frozen-in dark matter \cite{Chu:2011be,Forestell:2018dnu} through $e'\bar{e'} \rightarrow e'\bar{e'}e'\bar{e'}$ is negligible here,\footnote{To be conservative, we calculate the rate assuming all interactions take place at the maximum $\sqrt{s} \simeq m_\gamp$ and find that it is still far below Hubble. We perform the calculation of the cross section using MadGraph \cite{Alwall:2014hca} with a model implemented in Feynrules \cite{Alloul:2013bka}.} as is the depletion from $e' \bar{e'} \to \nu' \bar{\nu'}$.

An equal number of twin positrons are produced as twin electrons from the freeze-in processes. Requiring that $\epsilon$ reproduce the observed DM abundance today, we find
\al{
\epsilon = \sqrt{\frac{\Omega_\chi h^2 \rho_{\text{crit}}/h^2}{2m_\ep \Tilde{Y}_\ep s_0}},
}
where $\Omega_\chi h^2 \approx 0.12$, $\rho_{\text{crit}}/h^2 \approx 1.1 \times 10^{-5} \text{GeV}/\text{cm}^3$, and $s_0 \approx 2900/\text{cm}^3$ \cite{Tanabashi:2018oca}. $\Tilde{Y}_\ep$ is the total yield with the overall factor of $\epsilon^2$ removed.
\begin{figure}[t!]
\includegraphics[width = \columnwidth]{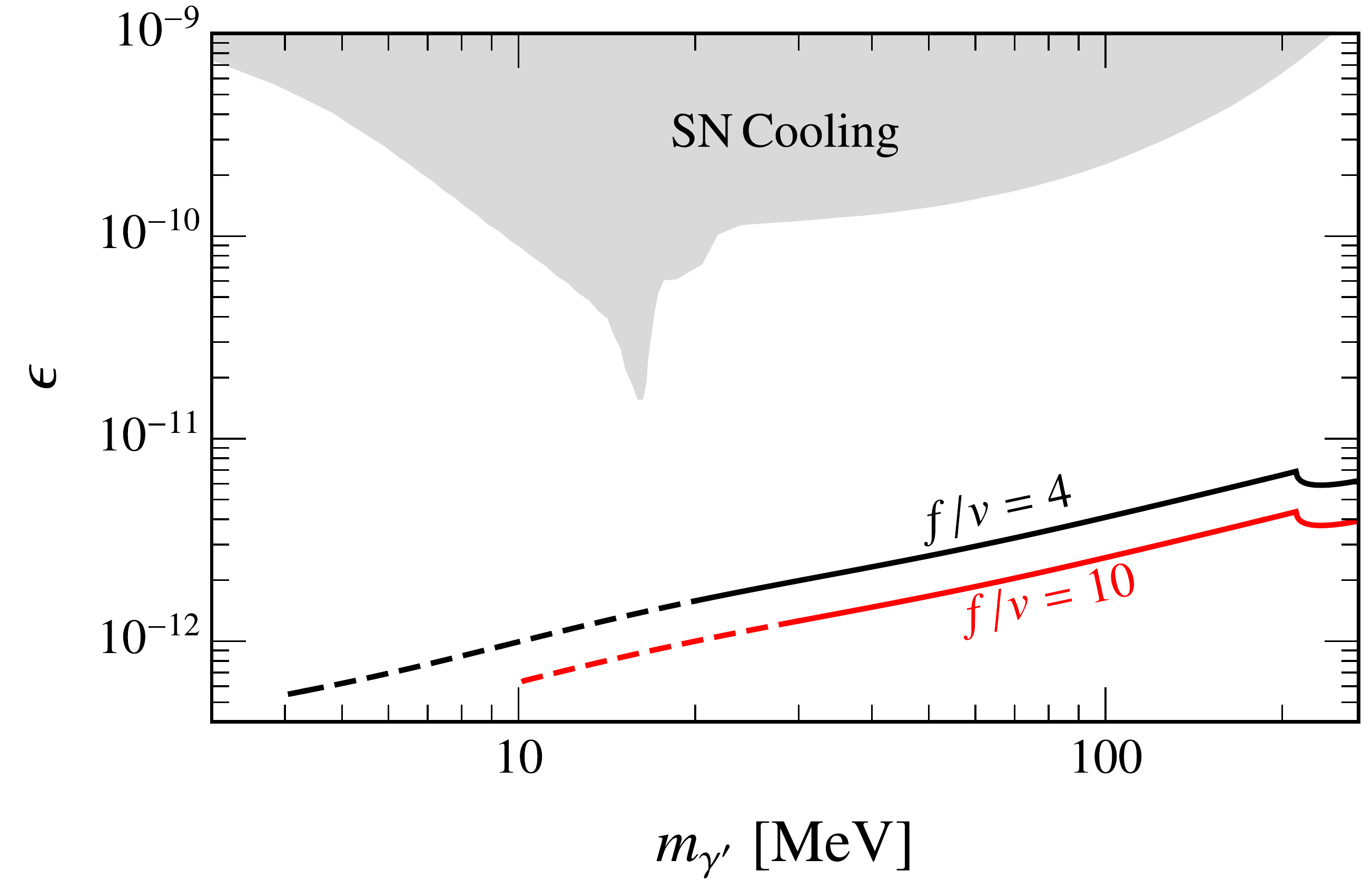}
\caption{\label{fig:mAvseps} Contours in the plane of twin photon mass $m_\gamp$ and kinetic mixing $\epsilon$ which freeze-in the observed DM abundance for two values of $f/v$. The dip at high masses corresponds to additional production via muon annihilations. In the dashed segments, self-interactions occur with $\sigma_{\text{elastic}}/m_{e'}\gtrsim 1 \text{cm}^2/\text{g}$. Also included are the combined supernova cooling bounds from \cite{Chang:2016ntp,Hardy:2016kme}.} 
\end{figure}
This requisite kinetic mixing appears in Fig.~\ref{fig:mAvseps} as a function of the twin photon mass $m_{\gamma'}$ for the two benchmark $f/v$ values 4 and 10. In grey, we plot constraints from anomalous supernova cooling. To be conservative, we include both, slightly different bounds from \cite{Chang:2016ntp,Hardy:2016kme}. The dashed regions of the lines show approximately where self-interactions through Bhabha scattering are relevant in the late universe, $\sigma_{\text{elastic}}/m_{e'}\gtrsim 1 \text{ cm}^2/\text{g}$. Self-interactions much larger than this are constrained by the Bullet Cluster~\cite{Clowe:2003tk,Markevitch:2003at,Randall:2007ph} among other observations. Interestingly, self-interactions of this order have been suggested to fix small-scale issues, and some claimed detections have been made as well. We refer the reader to \cite{Tulin:2017ara} for a recent review of these issues. 

As mentioned above and discussed further in Appendix \ref{sec:KinMixapp}, the level of kinetic mixing required for freeze-in is roughly of the same order as is expected from infrared contributions in the MTH. It would be interesting to develop the technology to calculate the high-loop-order diagrams at which it may be generated. In the context of a complete model of the MTH where kinetic mixing is absent in the UV, $\epsilon$ is fully calculable and depends solely on the scale at which kinetic mixing is first allowed by the symmetries. Calculating $\epsilon$ would then predict a minimal model at some $m_{\gamma'}$ to achieve the right dark matter relic abundance, making this effectively a zero-parameter extension of MTH models with asymmetric reheating. Importantly, even if $\epsilon$ is above those shown in Fig.~\ref{fig:mAvseps}, that would simply point to a larger value of $m_\gamp$ which would suggest that the parameter point depends in more detail on the mechanism of asymmetric reheating. We note that in the case that the infrared contributions to $\epsilon$ are below those needed here, the required kinetic mixing may instead be provided by UV contributions and the scenario is unaffected.

\section{Conclusion} \label{sec:conc}

The mirror twin Higgs is perhaps the simplest avatar of the Neutral Naturalness program, which aims to address the increasingly severe little hierarchy problem. Understanding a consistent cosmological history for this model is therefore 
crucial, and an important step was taken in \cite{Craig:2016lyx,Chacko:2016hvu}. As opposed to prior work, the cosmology of the MTH was remedied without hard breaking of the $\mathbb{Z}_2$ symmetry by utilizing asymmetric reheating to dilute the twin energy density. Keeping the $\mathbb{Z}_2$ as a good symmetry should simplify the task of 
writing high energy completions of these theories, but low-scale reheating may slightly complicate cosmology at early times. These works left as open questions how to set up cosmological epochs such as dark matter generation and baryogenesis in such models. We have here found that at least one of these questions has a natural answer. 

In this work, we have shown that twin electrons and positrons may be frozen-in as dark matter following asymmetric reheating in twin Higgs models. This requires extending the mirror twin Higgs minimally with a single free parameter: the twin photon mass. Freezing-in the observed DM abundance pins the required kinetic mixing to a level expected from infrared contributions in MTH models. In fact, the prospect of calculating the kinetic mixing generated in the MTH could make this an effectively parameter-free extension of the MTH. Compared to generic freeze-in scenarios, it is interesting in this case that the ``just so'' stories of feeble coupling and negligible initial density were already present for reasons entirely orthogonal to dark matter.

This minimalism in freeze-twin dark matter correlates disparate signals which would allow this model to be triangulated with relatively few indirect hints of new physics. If deviations in Higgs couplings are observed at the HL-LHC or a future lepton collider, this would determine $f/v$ \cite{Ahmed:2017psb,Chacko:2017xpd,Alipour-Fard:2018lsf,Alipour-fard:2018mre}, which would set the dark matter mass. An observation of anomalous cooling of a future supernova through the measurement of the neutrino `light curve' might allow us to directly probe the $m_{\gamma'},\epsilon$ curve \cite{Chang:2016ntp,Hardy:2016kme}, though this would rely on an improved understanding of the SM prediction for neutrino production.\footnote{We thank Jae Hyeok Chang for a discussion on this point.} Further astrophysical evidence of dark matter self-interactions would point to a combination of $f/v$ and $m_{\gamma'}$. All of this complementarity underscores the value of a robust experimental particle physics program whereby new physics is pursued via every imaginable channel. 

\section*{Acknowledgements}

We are grateful to the organizers of the TASI 2018 summer school, during which this work was initiated. We have benefited from many stimulating discussions with the other participants, especially Jae Hyeok Chang and Lindsay Forestell. We are also grateful for helpful conversations with Tim Cohen, Nathaniel Craig, Jeff Dror, Isabel Garcia-Garcia, Roni Harnik, Matthew McCullough, and Dave Sutherland. We thank Nathaniel Craig, Jeff Dror, Lindsay Forestell, and Matthew McCullough for comments on the manuscript.

The research of SK is supported in part by the US Department of Energy under the grant DE-SC0014129. RM was supported by the National Science Foundation Graduate Research Fellowship Program for a portion of this work. 

\textit{Note added:} Following the completion of this work, \cite{Dunsky:2019zqr} appeared on the arXiv. Freeze-in of heavy mirror electrons is discussed in a ``Mirror Higgs Parity'' framework, but there is little overlap with this work.

\appendix

\section{Kinetic Mixing in the MTH}
\label{sec:KinMixapp}

Since kinetic mixing plays a central role in freeze-twin dark matter, we discuss here at some length the order at which it is expected in the low-energy EFT. Of course, there may always be UV contributions which set $\epsilon$ to the value needed for freeze-in. However, if the UV completion of the MTH disallows such terms - for example, via supersymmetry, an absence of fields charged under both sectors, and eventually grand unification in each sector (see e.g. \cite{Berezhiani:2005ek,Falkowski:2006qq,Chang:2006ra,Craig:2013fga,Katz:2016wtw,Badziak:2017syq})- then the natural expectation is for mixing of order these irreducible IR contributions.

To be concrete, we imagine that $\epsilon = 0$ at the UV cutoff of the MTH, $\Lambda \lesssim 4 \pi f$. To find the kinetic mixing in the regime of relevance, at momenta $\mu \lesssim 1 \text{ GeV}$, we must run down to this scale. As we do not have the technology to easily calculate high-loop-order diagrams, our analysis is limited to whether we can prove diagrams at some loop order are vanishing or finite, and so do not generate mixing. Thus our conclusions are strictly always `we know no argument that kinetic mixing of this order is not generated', and there is always the possibility that further hidden cancellations appear. With that caveat divulged, we proceed and consider diagrammatic arguments in both the unbroken and broken phases of electroweak symmetry.

Starting in the unbroken phase, we compute the mixing between the hypercharge gauge bosons. Two- and three-loop diagrams with Higgs loops containing one gauge vertex and one quartic insertion vanish. By charge conjugation in scalar QED, the three-leg amplitude of a gauge boson and a complex scalar pair must be antisymmetric under exchange of the scalars. However, the quartic coupling of the external legs ensures that their momenta enter symmetrically. As this holds off-shell, the presence of a loop which looks like
\begin{center}
\includegraphics[width=0.28\textwidth]{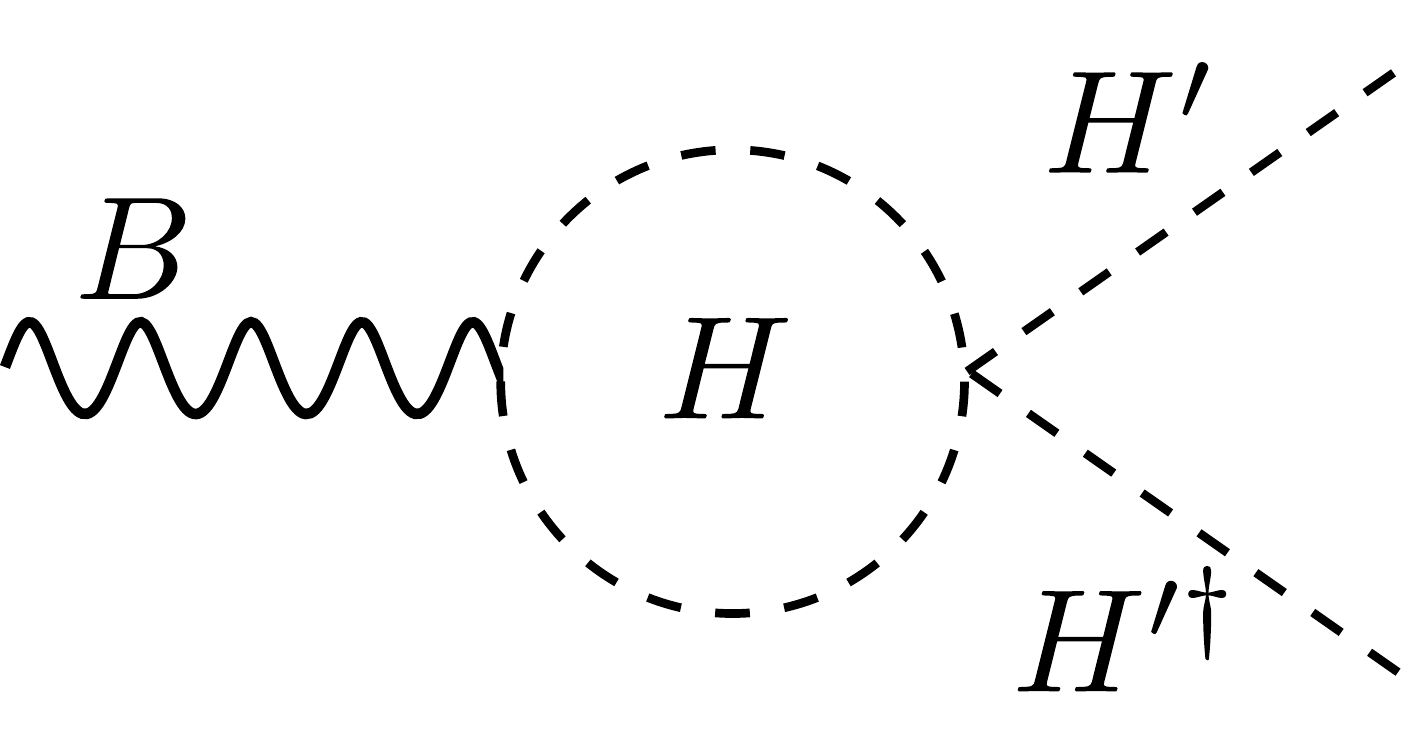}
\end{center}
causes the diagram to vanish. However, at four loops the following diagram can be drawn which avoids this issue:
\begin{center}
\includegraphics[width=0.45\textwidth]{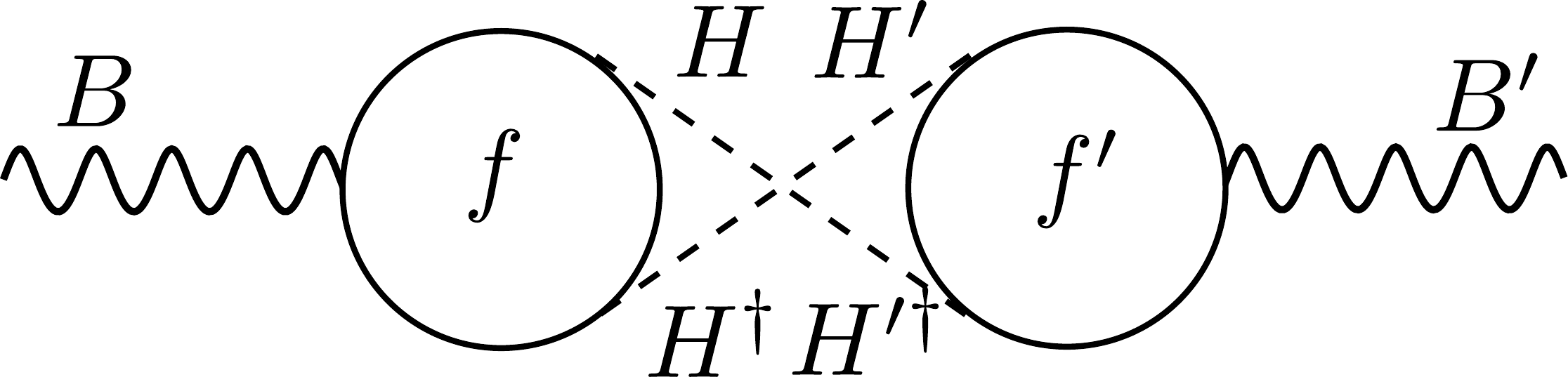}
\end{center}
where the two hypercharges are connected by charged fermion loops in their respective sectors and the Higgs doublets' quartic interaction. This diagram contributes at least from the MTH cutoff $\Lambda \lesssim 4 \pi f$ down to $f$, the scale at which twin and electroweak symmetries are broken. We have no argument that this vanishes nor that its unitarity cuts vanish. We thus expect a contribution to kinetic mixing of $\epsilon \sim g_1^2 c_W^2 / (4 \pi)^8$, with $g_1$ the twin and SM hypercharge coupling and $c_W = \cos \theta_W$ appearing as the contribution to the photon mixing operator. In this estimate we have omitted any logarithmic dependence on mass scales, as it is subleading.

In the broken phase, we find it easiest to perform this analysis in unitary gauge. The Higgs radial modes now mass-mix, but the emergent charge conjugation symmetries in the two QED sectors allow us to argue vanishing to higher-loop order. 
The implications of the formal statement of charge conjugation symmetry are subtle because we have two QED sectors, so
whether charge conjugation violation is required in both sectors seems unclear. However, similarly to the above case, there is a symmetry argument which holds off-shell. The result we rely on here is that in a vector-like gauge theory, diagrams with any fermion loops with an odd number of gauge bosons cancel pairwise. Thus, each fermion loop must be sensitive to the chiral nature of the theory, so the first non-vanishing contribution is at five loops as in:
\begin{center}
\includegraphics[width=0.45\textwidth]{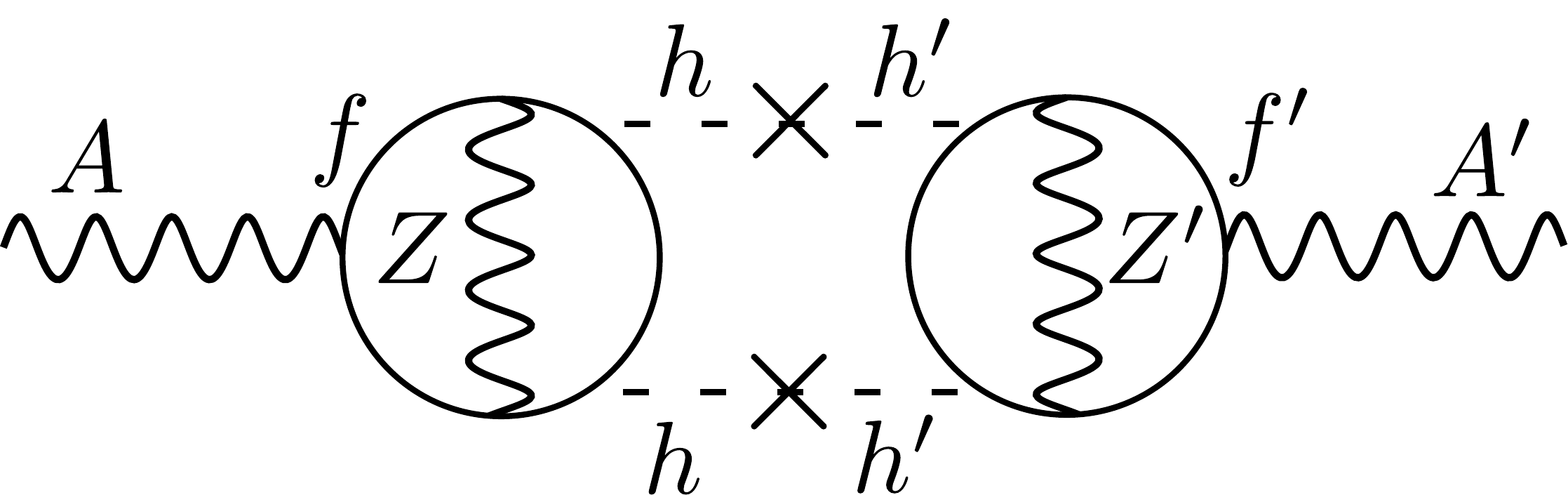}
\end{center}
where the crosses indicate mass-mixing insertions between the two Higgs radial modes which each contribute $\sim v/f$. Thus, both the running down to low energies and the finite contributions are five-loop suppressed. From such diagrams, one expects a contribution $\epsilon \sim e^2 g_A^2 g_V^2 (v/f)^2 / (4 \pi)^{10}$, where with $g_V$ and $g_A$ we denote the vector and axial-vector couplings of the $Z$, respectively. We note there are other five loop diagrams in which Higgses couple to massive vectors which are of similar size or smaller.

Depending on the relative sizes of these contributions, one then naturally expects kinetic mixing of order $\epsilon \sim 10^{-13} - 10^{-10}$. If $\epsilon$ is indeed generated at these loop-levels, then mixing on the smaller end of this range likely requires that it becomes disallowed not far above the scale $f$. However, we note that our ability to argue for higher-loop order vanishing in the broken versus unbroken phase is suggestive of the possibility that there may be further cancellations. We note also the possibility that these diagrams, even if nonzero, generate only higher-dimensional operators. Further investigation of the generation of kinetic mixing through a scalar portal is certainly warranted.

\bibliographystyle{kp}
\bibliography{refs}
\end{document}